# Simple Sensitivity Analysis for Differential Measurement Error

Tyler J. VanderWeele and Yige Li

Correspondence to: Tyler J. VanderWeele, Department of Epidemiology, Harvard T.H. Chan School of Public Health, 677 Huntington Avenue, Boston MA, Boston MA 02115 (tvanderw@hsph.harvard.edu)

Author affiliations: Department of Epidemiology, Harvard T.H. Chan School of Public Health, 677 Huntington Avenue, Boston MA, Boston MA 02115 (Tyler J. VanderWeele, and Yige Li)

Funding: The research was funded by NIH grant R01CA222147.

Conflict of interest: None declared.

Running head: Differential Measurement Error.

**Abstract**

Sensitivity analysis results are given for differential measurement error of either the exposure or outcome. In the case of differential measurement error of the outcome it is shown that the true effect of the exposure on the outcome on the risk ratio scale must be at least as large as the observed association between the exposure and the mis-measured outcome divided by the maximum strength of differential measurement error. This maximum strength of differential measurement error is itself assessed as the risk ratio of the controlled direct effect of the exposure on mis-measured outcome not through the true outcome. In the case of differential measurement error of the exposure, under certain assumptions concerning classification probabilities, the true effect on the odds ratio scale of the exposure on the outcome must be at least as large as the observed odds ratio between the mis-measured exposure and the outcome divided by the maximum odds ratio of the effect of the outcome on mis-measured exposure conditional on the true exposure. The results can be immediately used to indicate the minimum strength of differential measurement error that would be needed to explain away an observed association between an exposure measurement and an outcome measurement.

**Introduction**

Measurement error, along with unmeasured confounding and selection bias, are often considered as being the central threats to validity in observational studies (1-16). However, measurement error is perhaps more often dismissed as a threat than is bias due to potential unmeasured confounding. Some of this discrepancy may arise from the perception that at least non-differential measurement error often, though not always, biases results towards the null (1-6). In such contexts, when biases are towards the null, then when the original estimate is meaningfully large, the true effect will be even larger. While this intuition does not always hold, it is still used with some frequency as justification for ignoring questions of measurement error. However, such ignoring of measurement error can fail, and dramatically so, when measurement error is differential such that the measurement error in the outcome depends on the value of the exposure, or when the measurement error in the exposure in fact depends on the value of the outcome, as may arise in retrospective reporting of the exposure. While various techniques to address differential measurement have been put forward (6-11), many of these techniques require detailed knowledge on numerous parameters related to conditional sensitivities and specificities for misclassification. These can be difficult to obtain data on, or to specify or interpret in a sensitivity analysis, unless a validation study, with access to the gold-standard measurement, is carried out.

In this paper we address these issues surrounding *differential* measurement error by proposing simple sensitivity analysis techniques that assess how strongly the differential measurement error would have to be to completely explain away an observed exposure-outcome association. The parameters in the results given are relatively easy to interpret. They essentially correspond, in the case of differential measurement error of the outcome, to the effect on the risk ratio scale of the exposure on the outcome measurement conditional on the true outcome. In the case of differential measurement error of the exposure, they correspond, under certain assumptions about classification probabilities, to the effect on the odds ratio scale of the outcome on the exposure measurement conditional on the true exposure. Analogous results are available for the measurement error of continuous exposures and outcomes as well. Before moving to formal notation and the results, we begin with a motivating example, which we will return to after the methods have been presented.

**Motivating Example**

Zhong et al. (17) use data from 2,279 mothers in Peru to examine the effects of social support during pregnancy on ante-partum depression. Social support was assessed in these studies either by satisfaction with social support or with self-report of the number of support providers available. We will focus here on the analyses in which social support was operationalized by satisfaction with social support. Conditional on various social, demographic, and health characteristics, and having high social support at baseline prior to pregnancy, those who had low social support during pregnancy were at 1.51 higher odds (95% CI: 1.03, 2.22) of ante-partum depression than those with high social support during pregnancy. The social support exposure in this study was assessed by self-report with reference to the period during early pregnancy or since becoming pregnant; however self-reported social support was itself assessed at the same time as ante-partum depression. If depression itself alters self-reported social support, then this could potentially account for the whole of the association. In other words, differential measurement error of the social support exposure might potentially explain away the association. We will return to this example after presenting the methods.

**Definitions and Notation**

We will give results here for a binary outcome; results for a continuous outcome, which can be derived from the other research literature (12), are given in the Appendix. Let A denote a binary exposure and Y denote a binary outcome. Let C denote a set of measured covariates. Let A* and Y* denote respectively the measurements of A and Y that may be subject to measurement error. We will first consider the setting of measurement error with respect to the outcome Y and then will turn to the case of measurement error with respect to the exposure A. As our focus in this paper is on measurement error, we will assume that the measured covariates C suffice to control for confounding for the effect of true exposure A on the true outcome Y. The results given below are stated in terms of probabilities and are thus applicable even if this no unmeasured confounding assumption does not hold, but to interpret the various expressions below as the effects of the exposure on the outcome, the measured covariates C would have to suffice to control for confounding. Some similar simple sensitivity analysis results for unmeasured confounding have been provided previously (13,14).

We will say that the measurement error for the outcome Y is non-differential with respect to exposure A if the distribution of Y* conditional on Y and C is independent of A which we can denote in terms of probability distributions by $P(y^*|a,y,c) = P(y^*|y,c)$. In other words, the measurement error of Y is non-differential with respect to exposure A if, conditional on measured covariates C, the exposure A gives no information about the measurement Y* beyond that in the true value of Y. If this is not the case so that Y* in fact depends on exposure A even conditional on (Y,C) then we say that the measurement error is differential. Non-differential measurement error of the outcome is represented in the diagram in Figure 1A and differential measurement of the outcome is represented in Figure 1B (15,16). If the exposure A affects the measurement Y* through pathways other than through its effect on the true outcome Y, then measurement error will be differential. It is this context which, as noted above, can often induce substantial bias away from the null, that will be the focus of the present paper.

In the context of differential measurement error of the outcome we will let $p_a = P(Y=1|A=a,C=c)$ denote the probability of the true outcome conditional on the exposure A=a and we will let $p_a^* = P(Y^*=1|A=a,C=c)$ denote the probability of the outcome measurement Y*=1 conditional on the exposure A=a. The true risk ratio for the effect of the exposure on the outcome is then $p_1/p_0 = P(Y=1|A=1,C=c)/P(Y=1|A=0,C=c)$ and the risk ratio obtained with the mismeasured data is then denoted by $p_1^*/p_0^* = P(Y^*=1|A=1,C=c)/P(Y^*=1|A=0,C=c)$. The sensitivity analysis results will relate the true risk ratio to the risk ratio obtained with the mismesaured data. In the expressions, $p_1/p_0$ and $p_1^*/p_0^*$, above we have suppressed the conditioning on C=c in the subscripts. The notation and results that follow are all applicable with analyses conditional on the measured covariates C. However, for notational simplicity, we have omitted "c" from the subscripts.

**Simple Sensitivity Analysis for Differential Measurement Error of the Outcome**

In the context of differential measurement error of the outcome let $s_a$ denote the sensitivity of the measurement Y* for Y conditional on A=a and C=c, that is, the probability, conditional on A=a and C=c that when Y=1, we also have that Y*=1. In terms of probabilities, $s_a=P(Y^*=1|Y=1,A=a,C=c)$. Note that with differential measurement error the sensitivity of the measurement Y* can vary with the value of the exposure A; the sensitivity may for example be lower for the exposed than for the unexposed. Also let $f_a$ denote the false positive probability of the measurement Y* for Y conditional on A=a and C=c, that is, the probability, conditional on A=a and C=c that when in fact Y=0, we have the measurement that Y*=1. In terms of probabilities, $f_a=P(Y^*=1|Y=0,A=a,C=c)$. The false positive probability is also just 1 minus the specificity. Again with differential measurement error, the false positive probabilities of the measurement Y* can vary with the value of the exposure A.

Consider now the ratio of the sensitivity parameter conditional on the exposure being present versus absent i.e. $s_1/s_0= P(Y^*=1|Y=1,A=1,C=c)/P(Y^*=1|Y=1,A=0,C=c)$.

This ratio is essentially the risk ratio for the effect of the exposure A on the measurement Y* conditional on Y=1 and C=c. It corresponds to the effect represented by arrow from A to Y* in Figure 1B when Y=1. It assesses how much more likely the measurement of Y* is reported to be present, Y*=1, when the exposure is present versus absent, in settings in which the true outcome is present. Likewise, consider the ratio of the false positive probabilities $f_1/f_0$=P(Y*=1|Y=0,A=1,C=c)/P(Y*=1|Y=0,A=0,C=c). This is essentially the risk ratio for the effect of the exposure A on the measurement Y* conditional on Y=0 and C=c. It corresponds to the effect represented by arrow from A to Y* in Figure 1B when Y=0. It assesses how much more likely the measurement of Y* is reported to be present, Y*=1, when the exposure is in fact present versus absent, in settings in which the true outcome is in fact absent.

We can now state our first sensitivity analysis result.

Theorem 1. Under differential measurement error of the outcome, if $p_1/p_0 \geq 1$ then $p_1/p_0 \geq (p_1^*/p_0^*) / \max(s_1/s_0, f_1/f_0)$; and if $p_1/p_0 \leq 1$ then $p_1/p_0 \leq (p_1^*/p_0^*) / \min(s_1/s_0, f_1/f_0)$.

The proof of this and other results are given in the Web Appendix. Consider the case of a causative exposure so that the true risk ratio is greater than or equal to 1. The result states that the true effect of the exposure on the outcome on the risk ratio scale, $p_1/p_0$, must be at least as large as the observed association between exposure A and measurement Y* on the risk ratio scale, $p_1^*/p_0^*$, divided by the "maximum direct effect" of A on Y* not through Y. This "maximum direct effect" is itself given by the maximum of the sensitivity risk ratio for A on Y*, $s_1/s_0$, and the false positive probability ratio for the effect of A on Y*, $f_1/f_0$. This also implies that for the observed association between A and Y* to be completely explained away (i.e. reduced to the null) by differential measurement error so there is no true causative effect, then the strength of the differential measurement error, assessed as the direct effect risk ratio of the effect of A on Y* not through Y must be at least as large as the observed association between A and Y* for either when Y=1, which is $s_1/s_0$, or when Y=0, which is $f_1/f_0$. This is stated formally is Corollary 1.

Corollary 1. Under differential measurement error of the outcome, if $p_1^*/p_0^*>1$ then for the true association to in fact be null so that $p_1/p_0=1$, it must be the case that either $s_1/s_0 \geq p_1^*/p_0^*$ or that $f_1/f_0 \geq p_1^*/p_0^*$; and if $p_1^*/p_0^*<1$ then for $p_1/p_0=1$, it must be the case that either $s_1/s_0 \leq p_1^*/p_0^*$ or $f_1/f_0 \leq p_1^*/p_0^*$.

While the proof of Theorem 1, from which Corollary 1 follows, is not entirely straightforward (see Appendix), the result in Corollary 1 is relatively intuitive graphically insofar as if the A-Y arrow in Figure 1B is to be absent and we observe an association between A and Y* then the direct effect arrow from A to Y* not through Y must be at least as large as our observed association between A and Y*. Similar logic can of course be used to assess the extent of measurement error needed to shift the estimate to any other value.

A very simple sensitivity analysis thus consists of specifying the maximum effect that the investigator believes is possible for A to have on Y* independent of Y i.e. how much more likely is the outcome reported to be present, Y*=1, comparing the exposure present to absent, when the true outcome is present, or how much more likely when the true outcome is absent. Once this is specified one simply divides the observed association between A and Y* by this parameter corresponding to the maximum direct effect of A on Y* independent of Y; the true effect of A on Y must be at least this large. Alternatively, one might just report that for differential measurement error to completely explain away the effect of the observed association between exposure A and measurement Y* the magnitude of differential measurement error, assessed by the maximum direct effect of A on Y* not through Y, must be at least as large as the observed association between A and Y*. We will now consider similar results for differential measurement error of the exposure.

**Sensitivity Analysis for Differential Measurement Error of the Exposure**

Consider measurement error in the exposure A. We will say that the measurement error for the exposure A is non-differential with respect to outcome Y if the distribution of A* conditional on A and C is independent of Y which we can denote in terms of probability distributions by $P(a*|y,a,c) = P(a*|a,c)$. In other words, the measurement error of exposure A is non-differential with respect to outcome Y if, conditional on measured covariates C, the outcome Y gives no information about the measurement A* beyond that in the true value of A. If this is not the case so that A* in fact depends on Y even conditional on (A,C) then we say that the measurement error is differential. Non-differential measurement error of the exposure is represented in the diagram in Figure 2A and differential measurement of the exposure is represented in Figure 2B.[14,15] If the outcome Y affects the measurement A* through pathways other than through its being affected by true exposure A, then measurement error will be differential. We will now consider simple sensitivity analysis for differential measurement error of the exposure.

As before, we will let $p_a=P(Y=1|A=a,C=c)$ denote the probability of the true outcome conditional on the exposure A=a and we will now let $p_a'=P(Y=1|A^*=a,C=c)$ denote the probability of the outcome conditional on the measurement A*=a. To get tractable results for differential measurement error of the exposure, we will consider the odds ratio scale. The true odds ratio for the effect of the exposure on the outcome is then $\{p_1/(1-p_1)\}/\{p_0/(1-p_0)\}=\{P(Y=1|A=1,C=c)/P(Y=0|A=1,C=c)\}/\{P(Y=1|A=0,C=c)/P(Y=0|A=0,C=c)\}$ and the odds ratio obtain with the mismeasured data is then denoted by $\{p_1'/(1-p_1')\}/\{p_0'/(1-p_0')\}=\{P(Y=1|A^*=1,C=c)/P(Y=0|A^*=1,C=c)\}/\{P(Y=1|A^*=0,C=c)/P(Y=0|A^*=0,C=c)\}$. The sensitivity analysis results will relate the true odds ratio to the odds ratio obtained with the mismesaured data.

In the context of differential measurement error of the exposure let $s_y'$ denote the sensitivity of the measurement A* for A conditional on Y=y and C=c, that is, the probability, conditional on Y=y and C=c that when A=1, we also have that A*=1. In terms of probabilities, $s_y'=P(A^*=1|Y=y,A=1,C=c)$. Note that with differential measurement error the sensitivity of the measurement A* can vary with the value of the outcome Y. Also let $f_y'$ denote the false positive probability of the measurement A* for A conditional on Y=y and C=c, that is, the probability, conditional on Y=y and C=c that when in fact A=0, we have the measurement that A*=1. In terms of probabilities, $f_y'=P(A^*=1|Y=y,A=0,C=c)$.

Again with differential measurement error, the false positive probabilities of the measurement A* can vary with the value of the outcome Y.

Consider now the odds ratio related to the sensitivity of the measurement A* for A conditional on the outcome being present versus absent, $\{s_1'/(1-s_1')\}/\{s_0'/(1-s_0')\}=\{P(A^*=1|Y=1,A=1,C=c)/P(A^*=0|Y=1,A=1,C=c)\}/\{P(A^*=1|Y=0,A=1,C=c)/P(A^*=0|Y=0,A=1,C=c)\}$. This ratio is essentially the odds ratio for the effect of the outcome Y on the exposure measurement A* conditional on A=1 and C=c. It corresponds to the effect represented by arrow from Y to A* in Figure 2B when A=1. It assesses how higher the odds of the measurement of A* is reported to be present, A*=1, when the outcome is present versus absent, in settings in which the true exposure is in fact present. Likewise, consider the odds ratio of the false positive probabilities for the exposure, $\{f_1'/(1-f_1')\}/\{f_0'/(1-f_0')\}=\{P(A^*=1|Y=1,A=0,C=c)/P(A^*=0|Y=1,A=0,C=c)\}/\{P(A^*=1|Y=0,A=0,C=c)/P(A^*=0|Y=0,A=0,C=c)\}$. This is essentially the odds ratio for the effect of the outcome Y on the exposure measurement A* conditional on A=0 and C=c. It corresponds to the effect represented by arrow from Y to A* in Figure 2B when A=1. It assesses how higher the odds of the measurement of A* is reported to be present, A*=1, when the outcome is present versus absent, in settings in which the true exposure is in fact absent. In the context of the ante-partum depression example above, $\{s_1'/(1-s_1')\}/\{s_0'/(1-s_0')\}$ would be the odds ratio by which antepartum depression would increase the probability of self-reporting high social support for the group that did in fact have high social support; and $\{f_1'/(1-f_1')\}/\{f_0'/(1-f_0')\}$ would be the odds ratio by which antepartum depression would increase the probability of self-reporting high social support for the group that in fact had low social support.

For our second sensitivity analysis result, we also need to consider the ratio by which the presence of Y=1 leads to correct classification when A is present versus absent, and also the ratio by which the presence of Y=1 leads to incorrect classification when A is present versus absent. Define $r_c=\{s_1'/s_0'\}/\{(1-f_1')/(1-f_0')\}=\{P(A^*=1|Y=1,A=1,C=c)/P(A^*=1|Y=0,A=1,C=c)\}/\{P(A^*=0|Y=1,A=0,C=c)/P(A^*=0|Y=0,A=0,C=c)\}$ as the correct classification ratio. This is the ratio of how much more likely Y makes $A^*$ to be correctly classified, when A=1 versus when A=0. Define $r_i=\{f_1'/f_0'\}/\{(1-s_1')/(1-s_0')\}=\{P(A^*=1|Y=1,A=0,C=c)/P(A^*=1|Y=0,A=0,C=c)\}/\{P(A^*=0|Y=1,A=1,C=c)/P(A^*=0|Y=0,A=1,C=c)\}$ as the incorrect classification ratio. This is the ratio of how much more likely Y makes $A^*$ to be incorrectly classified, when A=1 versus when A=0. The result below will be of principal use when these correct and incorrect classification ratios are not larger than the extent of the differential measurement error direct effect of Y on $A^*$ captured by the odds ratios $\{s_1'/(1-s_1')\}/\{s_0'/(1-s_0')\}$ and $\{f_1'/(1-f_1')\}/\{f_0'/(1-f_0')\}$ above.

We can now state our second sensitivity analysis result. It essentially says that for an observed association between exposure measurement A* and outcome Y, the true effect of A on Y on the odds ratio scale must be at least as large as the observed association between A* and Y divided by the maximum strength of differential measurement error, assessed as the odds ratio of the effect of outcome Y on exposure measurement A* conditional on true exposure A and the correct and incorrect classification ratios. In terms of probabilities we have the following result.

Theorem 2. Under differential measurement error of the outcome, if $p_1/p_0 \geq 1$ then $\{p_1/(1-p_1)\}/\{p_0/(1-p_0)\} \geq \{p_1'/(1-p_1')\}/\{p_0'/(1-p_0')\} / \max[\{s_1'/(1-s_1')\}/\{s_0'/(1-s_0')\}, \{f_1'/(1-f_1')\}/\{f_0'/(1-f_0')\}, r_c, r_i]$; and if $p_1/p_0 \leq 1$ then $\{p_1/(1-p_1)\}/\{p_0/(1-p_0)\} \leq \{p_1'/(1-p_1')\}/\{p_0'/(1-p_0')\} / \min[\{s_1'/(1-s_1')\}/\{s_0'/(1-s_0')\}, \{f_1'/(1-f_1')\}/\{f_0'/(1-f_0')\}, r_c, r_i]$.

Consider the case of a causative exposure so that the true odds ratio is greater than or equal to 1. The theorem states that the true effect of the exposure on the outcome on the odds ratio scale, $\{p_1/(1-p_1)\}/\{p_0/(1-p_0)\}$, must be at least as large as the observed association between exposure A and measurement Y* on the odds ratio scale, $\{p_1'/(1-p_1')\}/\{p_0'/(1-p_0')\}$, divided by the maximum of the effect of Y on A* conditional on A i.e. the maximum of the sensitivity odds ratio for Y on A*, $\{s_1'/(1-s_1')\}/\{s_0'/(1-s_0')\}$, and the false positive probability odds ratio for the effect of Y on A*, $\{f_1'/(1-f_1')\}/\{f_0'/(1-f_0')\}$, provided the correct and incorrect classification ratios, $r_c$ and $r_i$, are not larger than the differential measurement error direct effects of Y on A* captured by the odds ratios $\{s_1'/(1-s_1')\}/\{s_0'/(1-s_0')\}$ and $\{f_1'/(1-f_1')\}/\{f_0'/(1-f_0')\}$. In this case, this also implies that for the observed association between A* and Y to be completely explained away (i.e. reduced to the null) by differential measurement error so there is no true causative effect, then the strength of the differential measurement error, assessed as the odds ratio for the effect of Y on A* conditional on A must be at least as large as the observed association between A* and Y for either when A=1, which is $\{s_1'/(1-s_1')\}/\{s_0'/(1-s_0')\}$, or when A=0, which is $\{f_1'/(1-f_1')\}/\{f_0'/(1-f_0')\}$. Note that if the outcome is rare, then Theorem 2 could also effectively be employed with risk ratios to compare $p_1/p_0$ to $p_1'/p_0'$ since the odds ratios for the outcome will then approximate risk ratios. However, even when this is the case, the sensitivity analysis odds ratio parameters must still be interpreted as odds ratios unless the exposure also is relatively rare. We will now illustrate Theorem 2 with an example.

**Example Revisited**

Recall Zhong et al. (17) use data from 2,279 mothers in Peru to examine the effects of social support during pregnancy on ante-partum depression. Conditional on various social, demographic, and health characteristics, and having high social support at baseline prior to pregnancy, those who had low social support during pregnancy were at 1.51 higher odds (95% CI: 1.03, 2.22) of ante-partum depression than those with high social support during pregnancy. There was concern that differential measurement error of the social support exposure might potentially explain away the association, since social support was reported retrospectively and assessed at the same time as ante-partum depression.

By Theorem 2 above, we would then have that for differential measurement error to be responsible for the whole of the observed association between social support and ante-partum depression, provided the correct and incorrect classification ratios were not larger than the differential measurement error direct effects of Y on A* as defined above, then the presence of antepartum depression would have to increase the odds of self-reporting high social support by at least 1.51-fold, either for the group that did in fact have high social support or for the group that in fact had low social support. This would constitute moderately strong differential measurement error but is perhaps not entirely implausible. We could examine other potential values as well; if we thought the effect would only be meaningful if it were at least as large as 1.1, we could state, under the assumptions above, that to reduce the estimated odds ratio from 1.51 to 1.1, the presence of antepartum depression would have to increase the odds of self-reporting high social support by at least 1.37-fold (=1.51/1.1), either for the group that did in fact have high social support or for the group that in fact had low social support. Note however, that while moderately strong differential measurement error would be needed to explain away the effect estimate, only very modest differential measurement error (e.g. differential measurement error direct effect odds ratios of 1.03) would be needed to shift the confidence interval to include 1.

**Discussion**

In this paper we have provided a relatively straightforward sensitivity analysis approach to evaluate the robustness or sensitivity of results to potential differential measurement error of the exposure or the outcome. The results led to the conclusion that the magnitude of the differential measurement error, quantified on the risk ratio scale for the direct effect of the exposure on the mis-measured outcome, or on the odds ratio scale for the effect of the outcome on the mis-measured exposure, must be at least as large as the corresponding observed ratios relating the exposure and outcome measurements. Like all sensitivity analysis techniques the results do not necessarily yield definite conclusions but can be helpful in establishing a general sense as to the robustness of results.

Of course, if data are available on the necessary sensitivity and specificities concerning the differentially mis-measured exposure or outcome then such data could be used to obtain more precise inferences about the magnitude of the true effect. However, data from such validation studies are often absent; it can be difficult to obtain such data even for a non-differentially mis-measured exposure or outcome, and for a differentially mis-measured exposure or outcome, it will be more challenging still. The results in this paper can be of use when such data are not available. While the present paper effectively uses these same sensitivity and specificity parameters, the results given here reparameterize these in an intuitive way so that it is easy to see how substantial differential measurement error would have to be to reduce the observed association to the null. In contrast, if one uses all four sensitivity and specificity parameters separately, there are numerous combinations of these which would together shift the observed estimate to the null so that the sensitivity analysis may be more difficult to interpret.

The results given here are thus particularly easy to use and can be employed without specifying the magnitude of the effects of differential measurement error simply by reporting (i) how strong the differential measurement error would need to be to completely explain away the estimate of the exposure-outcome association with the mis-measured variables and (ii) the minimum strength of differential measurement error needed to shift the confidence interval to include 1. An analogous approach, referred to as the "E-value" (13), has recently been put forward to address the minimum strength of unmeasured confounding necessary to explain away an observed exposure-outcome association. The E-value is defined as the "minimum strength of association on the risk ratio scale that an unmeasured confounder(s) would have to have with both the exposure and the outcome to explain away an observed exposure-outcome relationship." The E-value is always larger in magnitude than the magnitude of the risk ratio to be explained away. For example, a risk ratio of 1.1 yields an E-value of 1.43. In contrast, the magnitude of the minimum strength of differential measurement on the risk ratio or odds ratio scale needed to explain away an observed association is simply equal to the observed exposure-outcome measurement ratio. Thus, to explain away an observed risk ratio of 1.1 one only needs a differential measurement error for the outcome on the risk ratio scale of 1.1. Moreover, for differential misclassification of an exposure, the parameters were given on odds ratio scales and, since risk ratios are in general less extreme in magnitude than odds ratios (6,18), even less differential exposure differential exposure misclassification may be needed to explain away a given estimate.

It may seem from this analysis that measurement error is perhaps a stronger threat to the validity of etiologic studies than is unmeasured confounding. In the context of differential measurement error, that conclusion might often be warranted. We might then wonder why questions of measurement error seem to be addressed with somewhat less frequency than those of unmeasured confounding. The answer perhaps lies in part in the fact that, as noted in the introduction, non-differential measurement error often, though not always, biases estimates towards the null making them conservative. It is thus principally differential, rather than non-differential, measurement error, that typically has the potential for such large biases and substantial differential measurement error may be less common, at least outside of contexts in which exposures are reported retrospectively after the outcome occurs. Moreover, for many exposure and outcomes, measurement error, either differential or non-differential, may be minimal when more objective measurements are used, such as death or height, rather than those obtained by self-report. But this will of course vary across contexts and areas of research; even with mortality, measurement error may be present in registry data in developing countries for example. In contrast, in nearly any observational study, unmeasured confounding will likely be a concern. Exceptions can occur with natural experiments (and confounding for the exposure is of course eliminated on average in randomized trial) but these settings may be less frequent than those with measurements that are highly accurate and/or objective. The relatively greater emphasis on assessing the implications of unmeasured confounding over measurement error thus may not be entirely unreasonable, though both should arguably be assessed more often. And moreover, as shown here, when measurement error is differential, it can be particularly problematic insofar as the magnitude of effects needed to explain away an observed exposure-outcome association are considerably more modest for differential measurement error than they are for unmeasured confounding.

     We have focused here on differential measurement error of either a binary exposure or a binary outcome. However, analogous results are also possible for continuous exposure and/or outcome and can be obtained by adapting existing results in the literature (9,12). We state these formally in the Appendix. It is hoped that the results in this paper will allow for

a more straightforward and more frequent assessment of the potential of differential measurement to explain away, or substantially alter, effect estimates.

**Appendix. Simple Sensitivity Analysis for Differential Measurement Error of Continuous Outcomes and Exposures**

We first consider the case of differential measurement error of a continuous outcome in the context of linear regression. The proofs of these results and all results in the main text are given in the Web Appendix.

Theorem 3. If the measure $Y^*$ of outcome $Y$ is subject to differential measurement error with respect to exposure $A$ and if $E[Y^*|a,y,c] = \gamma_0 + \gamma_1 a + \gamma_2 y + \gamma_3'c$ and $E[Y|a,c] = \beta_0 + \beta_1 a + \beta_2'c$ and $E[Y^*|a,c] = \beta_0^* + \beta_1^* a + \beta_2^{*'}c$ then $\beta_1 = (\beta_1^* - \gamma_1)/\gamma_2$.

From Theorem 3 we have that if the outcome is subject to differential measurement error and we fit the regression model, $E[Y^*|a,c] = \beta_0^* + \beta_1^* a + \beta_2^{*'}c$ with the mis-measured outcome data then we can obtain a corrected estimate of the effect of A on the true outcome Y by $\beta_1 = (\beta_1^* - \gamma_1)/\gamma_2$ where $\gamma_1$ corresponds to the differential measurement error direct effect of A on $Y^*$ not through Y and where $\gamma_2$ corresponds to the effect of Y on $Y^*$. If $\gamma_2 = 1$ so that $Y^*$ scales with Y, then we have that $\beta_1 = \beta_1^* - \gamma_1$ so that we obtain the corrected estimate of the effect of A on Y simply by taking the estimate $\beta_1^*$ obtained using the observed data and subtracting from it $\gamma_1$, the differential measurement error direct effect of A on $Y^*$ not through Y.

We now consider an analogous result for differential measurement error of a continuous exposure.

Theorem 4. Suppose the measure A* of exposure A is subject to differential measurement error with respect to outcome Y and that $E[A^*|a,y,c] = \gamma_0 + a + \gamma_1 y + \gamma_2'c$. Let $\sigma^2_u$ denote the error variance in this regression, let $\sigma^2_a = Var(A|C)$ and let $\lambda = Var(A|C) / Var(A^*|C,Y) = \sigma^2_a/(\sigma^2_a+\sigma^2_u)$. For a continuous outcome Y with linear regressions $E[Y|a,c] = \beta_0 + \beta_1 a + \beta_2'c$ and $E[Y^*|a,c] = \beta_0^* + \beta_1^* a + \beta_2^{*'}c$, we have $\beta_1 = [\beta_1^* - \gamma_1/(\sigma^2_a+\sigma^2_u)] / \lambda$. For a rare binary outcome with logistic regression $logit(Y=1|a,c) = \theta_0 + \theta_1 a + \theta_2'c$ and $logit(Y^*=1|a,c) = \theta_0^* + \theta_1^* a + \theta_2^{*'}c$, we have $\theta_1 \approx [\theta_1^* - \gamma_1/(\sigma^2_a+\sigma^2_u)] / \lambda$.

From Theorem 4 we have that if the exposure is subject to differential measurement error and we fit the regression model $E[Y^*|a,c] = \beta_0^* + \beta_1^* a + \beta_2^{*'}c$ with the mis-measured exposure data then we can obtain a corrected estimate of the effect of true exposure A on outcome Y by $\beta_1 = [\beta_1^* - \gamma_1/(\sigma^2_a+\sigma^2_u)] / \lambda$, where we essentially subtract off a scaled version, $\gamma_1/(\sigma^2_a+\sigma^2_u)$, of the differential measurement error effect of Y on A*, from our estimate, $\beta_1^*$, of the mis-measured effect of A on Y, and then we rescale this again by diving by $\lambda$. The $\gamma_1/(\sigma^2_a+\sigma^2_u)$ that we subtract off from our the observed estimate, $\beta_1^*$, is essentially measured in standard deviations of A* (conditional on Y,C).

**Figure Legends**

Figure 1A. Non-differential measurement error of the outcome $Y$ as $Y^*$ with exposure $A$ and covariates $C$.

Figure 1B. Differential measurement error of the outcome $Y$ as $Y^*$ with exposure $A$ and covariates $C$.

Figure 2A. Non-differential measurement error of the exposure $A$ as $A^*$ with outcome $Y$ and covariates $C$.

Figure 2B. Differential measurement error of the exposure $A$ as $A^*$ with outcome $Y$ and covariates $C$.

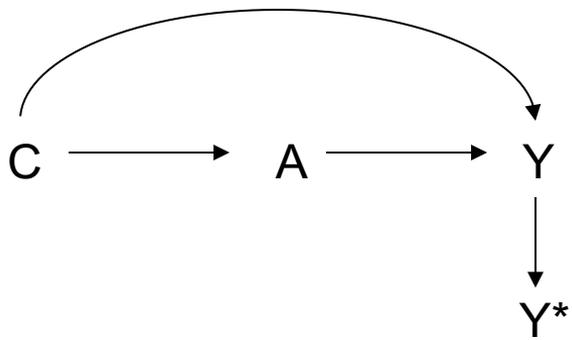

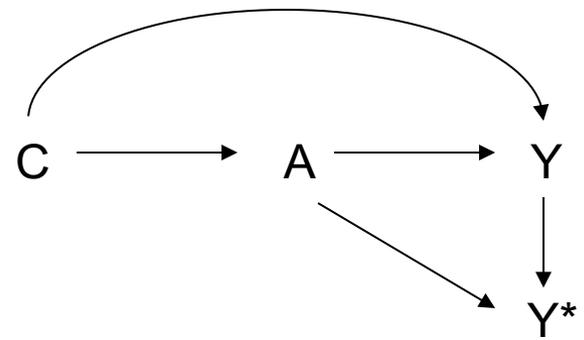

Fig. 1a. Non-differential measurement error of the outcome

Fig. 1b. Differential measurement error of the outcome

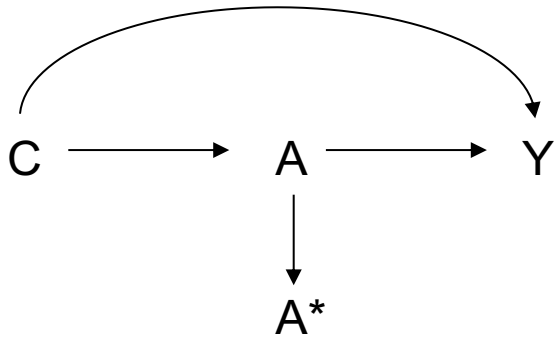

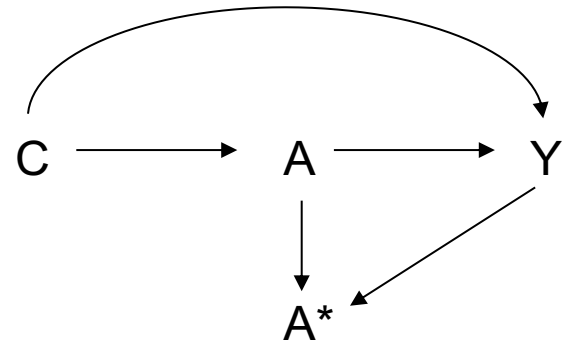

Fig. 2a. Non-differential measurement error of the exposure

Fig. 2b. Differential measurement error of the exposure